\documentclass[11pt,a4paper]{article}
\pdfoutput=1

\usepackage{jcappub}

\newcommand{\mpl}{{\rm M}_{\rm pl}}
\newcommand{\ksugra}{K_{\mathrm{sugra}}}
\newcommand{\wsugra}{W_{\mathrm{sugra}}}
\newcommand{\ksusy}{K_{\mathrm{susy}}}
\newcommand{\wsusy}{W_{\mathrm{susy}}}

\newcommand{\rD}{\ensuremath{\mathrm{D}}}

\newcommand{\ee}{\ensuremath{\mathrm{e}}}
\renewcommand{\bar}{\overline}

\newcommand{\kahler}{K\"{a}hler }

\def\cD{{\rD}}

\def\cO{{\cal O}}

\newcommand{\rar}{\rightarrow}

\newcommand{\non}{\nonumber\\}

\newcommand{\openone}{\leavevmode\hbox{\small1\normalsize\kern-.33em1}}

\title{
Decoupling limits in multi-sector supergravities
}

\subheader{\hfill {\rm preprint ITFA11-12}}

\author[a]{Ana Ach\'ucarro}
\author[a]{Sjoerd Hardeman}
\author[b]{Johannes M. Oberreuter}
\author[a]{Koenraad Schalm}
\author[a]{Ted van der Aalst}
\affiliation[a]{Instituut-Lorentz for Theoretical Physics, Universiteit Leiden,\\ \mbox{Niels Bohrweg 2, Leiden, {The Netherlands}}}
\affiliation[b]{Instituut voor Theoretische Fysica, Universiteit van Amsterdam,\\ \mbox{Science Park 904, Amsterdam, The Netherlands}}
\emailAdd{achucar@lorentz.leidenuniv.nl}
\emailAdd{j.m.oberreuter@uva.nl}
\emailAdd{kschalm@lorentz.leidenuniv.nl}
\emailAdd{vdaalst@lorentz.leidenuniv.nl}

\dedicated{Sadly, Sjoerd Hardeman passed away only a few days before the completion of this letter. \\ We wish to commemorate him as a splendid physicist and a good friend.}

\abstract{
Conventional approaches to cosmology in
  supergravity assume the existence of multiple
sectors that only communicate gravitationally.
In principle these
sectors decouple in the limit $\mpl\rightarrow \infty$. In
practice such a limit is delicate: for generic supergravities,
where sectors are combined by adding their \kahler functions, the
separate superpotentials must contain non-vanishing vacuum
expectation values {supplementing} the na\"{\i}ve global superpotential. We show that this
requires non-canonical scaling in the na\"{\i}ve supergravity
superpotential couplings to recover independent sectors of
globally supersymmetric field theory in the decoupling limit $\mpl
\to
\infty$.
}

\keywords{cosmology of theories beyond the SM, inflation, particle physics - cosmology connection, physics of the early universe, supersymmetry and cosmology}

\notoc

\begin{document}
\maketitle

\newpage
\section{Introduction}

Multiple sectors are a common feature in supergravity
cosmology and phenomenology.  These sectors are
  necessary to either incorporate inflation or supersymmetry breaking
  or are a consequence of string model-building. In particular to
study inflation, it is desirable to separate the dynamics of all
fields that do not contribute to the exponential expansion of the
Universe from the inflaton fields that do. Since gravity is the
weakest possible interaction, the inflationary sector is assumed to
only couple gravitationally to an unknown ``hidden''
sector that may also break supersymmetry by itself.  Whereas it is
natural for a rigid supersymmetric theory to be separated into several
sectors, the restrictive structure of supergravity forces the
different sectors to couple not only non-locally through graviton
exchange but also directly. For this reason embedding supersymmetric
theories as sectors into a supergravity can be notoriously difficult,
see e.g.
\cite{Nilles:1983ge,Wess:1992cp,Banks:1993en,Arnowitt:1993qp,Bagger:1994hh,Bagger:1994jh,Munoz:1995yp,Dine:2000bf}.

Though multiple sector supergravities are a long studied subject, the
context of cosmology has seriously sharpened the question. In
supergravity models of inflation it is commonly noted that
one seeks a consistent truncation of the scalar sector. This is
necessary but not sufficient. Even with a consistent truncation one
may have dominating instabilities towards the na\"{\i}vely non-dynamical
sectors, that can
 move them away from their supersymmetric critical points. One needs
either a symmetry constraint or an energy barrier to constrain the
dynamics to the putative inflaton sector.

Moreover, during inflation, supersymmetry is broken. Although it is frugal
to consider scenarios where the inflaton sector is also
responsible for phenomenological supersymmetry breaking (see e.g. \cite{AlvarezGaume:2010rt,Dine:2011ws,Achucarro:2012hg}), this need not
be so. 
For instance, in a generic gauge-mediation scenario, the mechanism responsible for supersymmetry breaking need not involve the fields that drive inflation.
This example immediately
  shows that the generic cosmological set-up must be able to account
  for a sector that breaks supersymmetry {\em independently} of the
  inflationary dynamics.

    This notion is our starting point. We consider a
    multiple-sector supergravity that decouples in the
    strictest sense
    in the limit $\mpl\to\infty$. In this limit the action must then
    be the sum of two independent functions\footnote{As example we
      consider the simplest case, a model with two uncharged scalar
      supermultiplets $X^a = (\phi, z)$ that are singlets under all
      symmetries. Gauge interactions and global symmetries will not
      change this general argument provided the two sectors are not
      mixed by symmetries or coupled by gauge fields. Therefore, we
      will also ignore $D$-terms in the supergravity potential below.}
\begin{equation}
   \label{eq:twosectoraction}
 S[\phi,\bar{\phi},z,\bar{z}]=S[\phi, \bar{\phi}]+S[z, \bar{z}] \;,
\end{equation}
such that the path integral factorizes. For a globally supersymmetric field theory with a standard kinetic term this can be achieved by demanding that the independent \kahler and superpotentials sum
\begin{align}\label{eq:defglobalsusy}
K_{\mathrm{susy}}(\phi,\bar{\phi},z,\bar{z}) &=
K^{(1)}(\phi,\bar{\phi}) + K^{(2)}(z,\bar{z}) \;, &
W_{\mathrm{susy}}(\phi,z) &= W^{(1)}(\phi) + W^{(2)}(z) \;.
\end{align}
The issue we address here is that
in supergravity complete decoupling in the sense of
(\ref{eq:twosectoraction}) appears to be impossible, even in
principle. Even with block diagonal kinetic terms from a sum of
\kahler potentials, the more complicated form of the supergravity
potential
\begin{equation}\label{eq:sugraV}
V_{\mathrm{sugra}} = \ee^{K/\mpl^2} \left(K^{a\bar{b}}\cD_aW\overline{\cD_{b}W} - \frac{3
|W|^2}{\mpl^2} \right)\;,~~ \cD \wsugra=\partial \wsugra+\partial
\ksugra\frac{\wsugra}{\mpl^2} \;,
\end{equation}
implies that there are many {\em direct} couplings between the two
sectors.  It raises the immediate question: if the low-energy
$\mpl\to\infty$ globally supersymmetric model must consist of
decoupled sectors, what is the relation between $K_{\mathrm{sugra}},
W_{\mathrm{sugra}}$ and $K_{\mathrm{susy}}, W_{\mathrm{susy}}$, or
vice versa given a globally supersymmetric model described by
$K_{\mathrm{susy}}, W_{\mathrm{susy}}$, what is the best choice for
$K_{\mathrm{sugra}}, W_{\mathrm{sugra}}$ such that the original theory
can be recovered in the limit $\mpl \to \infty$?

In this note we shall show that the scaling implied by the explicit factors of $\mpl$ in the
supergravity potential \eqref{eq:sugraV} is an incomplete answer to
this question. The direct communication between the sectors,
controlled by $\mpl$, has serious consequences for both the ground
state structure (solutions to the equation of motion,
i.e. the cosmological dynamics) and the interactions between the two
sectors. To be explicit, the first guess at how the rigid
supersymmetry and supergravity \kahler potentials and superpotentials
are related
\begin{equation}\label{eq:guesssugra}
\ksugra(\phi,\bar{\phi},z,\bar{z}) = \ksusy^{(1)}(\phi,\bar{\phi}) + \ksusy^{(2)}(z,\bar{z}) + \ldots
\;,~~
\wsugra(\phi,z) = \wsusy^{(1)}(\phi) + \wsusy^{(2)}(z) + \ldots \;,
\end{equation}
with $\ldots$ indicating Planck-suppressed terms and possibly a constant term, suffers from the
drawback that the ground states of the full theory are no longer
the product of the ground states of the individual sectors, except
when both (rather than only one) ground states are supersymmetric
\cite{deAlwis:2005tf,deAlwis:2005tg} (see also
\cite{Achucarro:2008sy,Achucarro:2007qa,Achucarro:2008fk}). This
directly follows from considering the extrema of the supergravity
potential\footnote{To derive (\ref{eq:Via}) note that, since $\cD
W/W$ is \kahler invariant and since the Levi-Civita
connection $\nabla$ of the field space manifold does not get
cross-contributions in a product manifold,
\begin{equation*}
\nabla_i\frac{\cD_\alpha W}{W}=
\partial_i\frac{\cD_\alpha W}{W}=
\cD_i\frac{\cD_\alpha W}{W}.
\end{equation*}}
\begin{align}
\nabla_i V&=\frac{\cD_i W}{W}V+\ee^{K/\mpl^2}|W|^2\left(\nabla_i\left(\frac{\cD_jW}{W}\right)\frac{\cD^j\bar{W}}{\bar{W}}+\frac{1}{\mpl^2}\frac{\cD_iW}{W}+\nabla_i\left(\frac{\cD_\beta W}{W}\right)\frac{\cD^\beta\bar{W}}{\bar{W}}\right) \;,\label{eq:Vi}\\
\nabla_i\nabla_\alpha V&=\frac{\cD_\alpha W}{W}\nabla_iV+\frac{\cD_iW}{W}\nabla_\alpha V-\frac{\cD_iW}{W}\frac{\cD_\alpha W}{W}V
+\cD_i\left(\frac{\cD_\alpha W}{W}\right)(V+\frac{2}{\mpl^2}\ee^{K/\mpl^2}|W|^2)\nonumber\\
&\phantom{=}
+\ee^{K/\mpl^2}|W|^2\left(\nabla_i\nabla_\alpha\left(\frac{\cD_\beta
W}{W}\right)\frac{\cD^\beta
\bar{W}}{\bar{W}}+\nabla_\alpha\nabla_i\left(\frac{\cD_j
W}{W}\right)\frac{\cD^j \bar{W}}{\bar{W}}\right) \label{eq:Via}\;.
\end{align}
Supersymmetric ground states, for which
the covariant derivatives of $W$ vanish on the solution,
$\cD_iW=0$ and $\cD_{\alpha}W=0$, are still product solutions. But for \kahler and superpotentials that sum \eqref{eq:guesssugra}, even if only one sector is in a non-supersymmetric ground state, by which we mean $\cD_iW=0$, $\cD_{\alpha}W\neq 0$, we can neither
conclude that sector 2, labeled by $i$, is in a minimum, for which $\nabla_i V$
would vanish, nor that the condition for sector 1, labeled by $\alpha$, to be in
a local ground state is independent of the
sector 2
fields $z^i$, which would
mean that $\nabla_i\nabla_\alpha V=0$. The former is only true
when
\begin{equation}\label{eq:Vicondition}
 \nabla_i\left(\frac{\cD_\beta W}{W}\right)\frac{\cD^\beta\bar{W}}{\bar{W}}=0 \;.
\end{equation}
The second requires, in addition,
\begin{equation}\label{eq:Viacondition}
\nabla_i\nabla_\alpha\left(\frac{\cD_\beta W}{W}\right)\frac{\cD^\beta \bar{W}}{\bar{W}}+\nabla_\alpha\nabla_i\left(\frac{\cD_j W}{W}\right)\frac{\cD^j \bar{W}}{\bar{W}}=0\;,
\end{equation}
and also sharpens the first condition \eqref{eq:Vicondition} to\footnote{These conditions
are merely sufficient not necessary. However, it is clear that the
restrictive nature of supergravity enforces conditions on the
unknown sectors for the system to be separate.}
\begin{equation}
\label{eq:Viacondition2}
\cD_i \frac{\cD_{\alpha} W}{W} =0 \;.
\end{equation}
Equations (\ref{eq:Vicondition}--\ref{eq:Viacondition2}) are
conditions for decoupling which apply not only to the ground state of
the full system but also to other critical points of the potential,
for instance along an inflationary valley. Generically these
conditions are not met on the solution (the second derivative need not
vanish at an extremum; recall that $\cD_iW$ does not vanish
identically but only on the solution).  Hence, generically the ground
states of hidden sectors mix and this spoils many cosmological
supergravity scenarios that truncate the action to one or the other
sector (see e.g. \cite{Gallego:2011jm} and references
therein). It is this issue that is particularly relevant
for inflationary model building, where a very weak coupling between
the inflaton sector and all other sectors has to persist over an
entire {\it trajectory} in field space where the expectation values of
the fields are changing with time (see e.g.
\cite{BenDayan:2008dv,Davis:2008sa,Achucarro:2010jv,Achucarro:2010da,Hardeman:2010fh}). At
  the same time, one is interested in the generic situation in which
  {\em both} sectors may contribute to supersymmetry breaking.\footnote{This situation has to be contrasted to phenomenological
    models appropriate for studying gravity mediated supersymmetry
    breaking, such as an ansatz \cite{Kaplunovsky:1993rd}
\begin{equation*}
 K(\phi,\bar{\phi},z,\bar{z}) =K_0(\phi,\bar{\phi})+z^i\bar{z}^{\bar{j}}K_{i\bar{j}}(\phi,\bar{\phi}) \;, \qquad W(\phi,z)=W_0(\phi)+z^iz^jW_{ij}(\phi)\;.
\end{equation*}
or equivalently, if $W \neq 0$,
$$G(\phi,\bar{\phi},z,\bar{z})=G^{(0)}(\phi,\bar{\phi})+z^i\bar{z}^{\bar{j}}
G^{(1,1)}_{i\bar{j}}(\phi,\bar{\phi})+z^iz^j
G^{(2,0)}_{ij}(\phi,\bar{\phi})+\bar{z}^{\bar{i}}\bar{z}^{\bar{j}}
G^{(0,2)}_{\bar{i}\bar{j}}(\phi,\bar{\phi})+\dots\;. $$
In models like these, it is understood that $\dot{z}=0$ and the
$z$-sector can remain in its supersymmetric critical point throughout
the evolution of the supersymmetry breaking fields.  For inflation,
such an expectation is unrealistic, as the supersymmetry-preserving
sector can become unstable during the inflationary dynamics, see
e.g. a recent discussion of the case in which the inflaton field
$\phi$ is solely responsible for supersymmetry breaking during
inflation (\cite{Achucarro:2012hg} and references therein). In this
relatively simple case, and except for very fine-tuned situations, the
generic scenario appears to be that one or more of the $z$-fields are
destabilized somewhere along the inflationary trajectory and they
trigger an exit from inflation (in other words, they become
``waterfall'' fields, and inflation is of the hybrid
kind \cite{Linde:1993cn}).  This implies that the pattern of
supersymmetry breaking today is not related to the one during
inflation, and also, since the waterfall fields
are forced away from their supersymmetric critical points,
 that supersymmetry is broken by both sectors as
the Universe evolves towards the current vacuum.}

\section{Natural multi-sector supergravities}
There is a well-known natural way to construct supergravity
potentials for which the ground states (and critical points) do
separate better. This obvious combination of superpotentials
automatically satisfies
(\ref{eq:Vicondition}--\ref{eq:Viacondition2}) and hence does
ensure that if one of the ground states is supersymmetric, the
ground state of the other sector is a decoupled field theory
ground state whether it breaks supersymmetry or not. This is if we
choose a product of superpotentials, keeping the sum of \kahler
potentials as before,
\begin{equation}\label{eq:defsugra}
  K_{\mathrm{sugra}}(\phi,\bar{\phi},z,\bar{z}) = K_{\mathrm{sugra}}^{(1)}(\phi,\bar{\phi}) + K_{\mathrm{sugra}}^{(2)}(z,\bar{z}) \;, \quad
  W_{\mathrm{sugra}}(\phi,z) = \frac{1}{\mpl^3} W_{\mathrm{sugra}}^{(1)}(\phi)W_{\mathrm{sugra}}^{(2)}(z) \;.
\end{equation}
This is well-known
\cite{Cremmer:1982vy,Binetruy:1984wy,Barbieri:1985wq} and has
recently been emphasized in the context of cosmology
\cite{Hsu:2003cy,Binetruy:2004hh,Achucarro:2007qa,Achucarro:2008sy,Achucarro:2008fk,Davis:2008sa,Achucarro:2010jv,Hardeman:2010fh,Gallego:2011jm}.
This ansatz conforms to the more natural description of
supergravities in terms of the K\"ahler invariant function
\begin{equation}
\label{eq:defG}
G(X,\bar{X}) =  \frac{1}{\mpl^2} \ksugra(X,\bar{X}) +
\log\left(\frac{\wsugra(X)}{\mpl^3}
\right) + \log\left( \frac{\bar{W}_{\mathrm{sugra}}(\bar{X})}{\mpl^3}
\right)\;,
\end{equation}
which can be defined if $W$ is non-zero in the region of
interest.\footnote{We expect
  this condition to hold around a supersymmetry breaking vacuum with
  almost vanishing cosmological constant. It also holds in many models
  of supergravity inflation, although a notable exception is
  \cite{Kallosh:2010ug,Kallosh:2010xz}.}
This K\"ahler function in turn underlies
a
better description of
multiple
sectors in supergravity
where $G$ is a sum of independent functions
\begin{equation}\label{eq:twosectorG}
G(\phi,\bar{\phi},z,\bar{z}) = G^{(1)}(\phi,\bar{\phi}) + G^{(2)}(z,\bar{z}) \;,
\end{equation}
such that the two sectors are separately K\"ahler invariant.  The sum
implies the product superpotential put forward above. This
  is the simplest ansatz that still  allows some degree of calculational
  control when both sectors break supersymmetry ---as well as
  optimizing decoupling along the inflationary trajectory. One of the
  simplest models of hybrid inflation in supergravity, $F$-term
  inflation \cite{Dvali:1994ms,Linde:1997sj}, is in this class.

\section{Decoupling}\label{sec:decoupling}
Given that we have just argued that a product of superpotentials is a
more natural framework to discuss multiple sector
supergravities, the obvious question arises how to recover a decoupled
{\em sum} of potentials for a globally supersymmetric theory in the
limit where gravity decouples, i.e. in which
\begin{equation}
V_{\mathrm{sugra}} = \ee^{K/\mpl^2} \left( |\cD W|^2 - \frac{3
|W|^2}{\mpl^2} \right)~~ \to ~~V_{\mathrm{susy}}=\sum_j|\partial_j W^{(j)}|^2 \;.
\end{equation}
For a two-sector supergravity defined by eqs. (\ref{eq:defsugra}) one
would not find this answer, if one takes the standard decoupling limit
$\mpl\rar \infty$ with both $K=K^{(1)}+K^{(2)}$ and
$W=\mpl^{-3}W^{(1)}W^{(2)}$ fixed.\footnote{Strictly speaking the
  decoupling limit sends $\mpl\rar\infty$ while keeping the fields
  $\phi, z$ fixed with $W^{(j)}/\mpl^3$ a holomorphic function of
  $\phi/\mpl$ or $z/\mpl$ and $K^{(j)}/\mpl^2$ a real function of
  $\phi/\mpl, \bar{\phi}/\mpl$ or $z/\mpl,\bar{z}/\mpl$. The limit
  zooms in to the origin so $K$ must be assumed to be non-singular
  there. Formally the decoupling limit does not exist
  otherwise. Physically it means that one is taking the decoupling
  limit w.r.t. an a priori determined ground state, around which $K$
  and $W$ are expanded. If $K$ is non-singular at the origin, the
  overall factor $\ee^{{K}/\mpl^2}$ yields an overall constant as
  $\mpl\rar \infty$, which may be set to unity, i.e. the constant part
  of $K$ vanishes. In the decoupling limit, both $K$ and $W$ may then
  be written as polynomials. Letting the coefficients in $W$ and $K$
  scale as their canonical scaling dimension such that $W$ has mass
  dimension three and $K$ has mass dimension two, then gives the rule
  of thumb that both $K$ and $W$ are held fixed as $\mpl \rar \infty$
\label{fn:1}
}
Instead, the product structure of the superpotential introduces a
cross-coupling between sectors,
\begin{equation}
  \label{eq:5}
  V_{\mathrm{eff}} = \frac{1}{\mpl^3} \left( |W^{(2)}|^2|\partial_\alpha W^{(1)}|^2+|W^{(1)}|^2|\partial_i W^{(2)}|^2 \right) \neq
  V_{\mathrm{susy}}\;,
\end{equation}
whose behavior under the limit $\mpl\to\infty$ is best examined at
the level of the superpotential.

Supergravity is sensitive to the expectation value $W_0=\langle
W\rangle$ of $W$, which
   relates the scale of supersymmetry breaking to the
expectation value of the potential, i.e.
 the cosmological constant
\begin{equation}
\Lambda^2 \mpl^2  = \langle V \rangle
 \sim \langle \cD W ^2\rangle - \frac{3}{\mpl^2} \langle W^2 \rangle
 = m_{\mathrm{susy}}^4 - 3 \frac{W_0^2}{\mpl^2}\;.
\end{equation}
The vacuum expectation value cannot vanish in a supersymmetry
breaking vacuum with (nearly) zero cosmological constant, such as
our Universe. Therefore, in the following we assume $\langle W \rangle \neq 0$ in
the region of interest. Instead of the usual way to incorporate
it, $W_{\mathrm{sugra}}=W_0+W_{\mathrm{dyn}}$ with
$W_{\mathrm{dyn}} = W_{\mathrm{susy}} +\ldots$, we include the vacuum expectation value for a two-sector
product superpotential by writing
\begin{align}\label{eq:productWs}
W(\phi,z)&=\frac{1}{\mpl^3}W^{(1)}W^{(2)} =\frac{1}{\mpl^3}
\left(W_0^{(1)}+W^{(1)}_{\mathrm{dyn}}(\phi)\right)
\left(W_0^{(2)}+W^{(2)}_{\mathrm{dyn}}(z)\right)\nonumber
\\&=\frac{1}{\mpl^3}\left( W_0^{(1)}W_0^{(2)} + W_0^{(2)}W^{(1)}_{\mathrm{dyn}}(\phi)+ W_0^{(1)}W^{(2)}_{\mathrm{dyn}}(z) + W^{(1)}_{\mathrm{dyn}}(\phi) W^{(2)}_{\mathrm{dyn}}(z)\right).
\end{align}
This is physically equivalent to a sum of superpotentials except
for the last term. Note again, that if one uses the standard
scaling, $\frac{\phi}{\mpl} \to 0$; $\frac{z}{\mpl} \to 0$ with
all couplings in $W^{(\mathrm{total})}$ having the canonical
scaling dimensions, this last term contains renormalizable
couplings involving the scalar partner of the goldstino, and these
are not Planck-suppressed: if supersymmetry is broken by the
$\phi$ sector, terms of the form ${\phi}{z}^2$ are renormalizable
and would survive the $\mpl
\to \infty$ limit, leading to a direct coupling between the two
sectors.\footnote{For a product of superpotentials we can always
choose a
\kahler gauge \emph{at every point} with $\langle K \rangle = \langle \partial_\phi K
\rangle =
\langle
\partial_z K \rangle = 0$ without mixing the superpotentials. In that case $F$-term supersymmetry
breaking is given by the linear terms in the expansion of
$W^{(1)}$ and $W^{(2)}$: $\langle \cD_\phi W \rangle  \sim \langle
\partial_\phi W^{(1)} \rangle$, $\langle \cD_z W \rangle \sim \langle
\partial_z W^{(2)} \rangle$.} If both sectors break supersymmetry
then mass-mixing terms ${\phi}{z}$ also survive. All such
(relevant) terms are of course absent if none of the two sectors
break supersymmetry, but this is not the case we are interested
in. One would have expected that these cross-couplings naturally
vanish in the decoupling limit.

The point of this note is simply to remark that the realization
that each of the superpotentials
$W^{(j)}=W^{(j)}_0+W_{\mathrm{dyn}}^{(j)}$ contains a constant
term can resolve this conundrum by assuming a non-standard scaling
for the constituent parts $W^{(j)}_0$, $W_{\mathrm{dyn}}^{(j)}$.
To achieve a decoupling we need that the cross term
$W_{\mathrm{dyn}}^{(1)}W_{\mathrm{dyn}}^{(2)}$, which contains the
coupling between the two sectors, scales away in the limit
$\mpl\rar\infty$. As a result the first term in
\eqref{eq:productWs} has to diverge, because its product with the
cross term should remain finite. In particular we can choose an
overall scaling
\begin{equation}\label{eq:twosectorscaling}
 W = \frac{1}{\mpl^{3}} ( \underbrace{W^{(1)}_0 W^{(2)}_0}_{\sim \mpl^{3+r}} + \underbrace{W^{(1)}_0 W^{(2)}_{\mathrm{dyn}}}_{\sim \mpl^3} + \underbrace{W^{(2)}_0 W^{(1)}_{\mathrm{dyn}}}_{\sim \mpl^3} + \underbrace{W^{(1)}_{\mathrm{dyn}} W^{(2)}_{\mathrm{dyn}}}_{\sim \mpl^{3-r}} ) \;,
\end{equation}
with $r>0$.
Let us account for dimensions by introducing an extra scale $m_\Lambda$ such that
\begin{align}
 \label{eq:scaling}
  W_{0}^{(1)} &= m_{\Lambda}^{\frac{3-r}{2} -A} \mpl^{\frac{3+r}{2}
  +A}\;,&
W_{\mathrm{dyn}}^{(1)} &=
\mpl^3\frac{W_{\mathrm{susy}}^{(1)}}{W_0^{(2)}}\;,\non W_{0}^{(2)}
&= m_{\Lambda}^{\frac{3-r}{2} +A} \mpl^{\frac{3+r}{2} -A}
\;,& W_{\mathrm{dyn}}^{(2)} &=\mpl^3
\frac{W_{\mathrm{susy}}^{(2)}}{W_0^{(1)}}\;,
\end{align}
with $W_{\mathrm{susy}}^{(j)}$ fixed as $\mpl \rar \infty$. Formally one can choose an inhomogeneous scaling with $A\neq 0$, but as we shall see it has no real consequences. For any $A$
it is easily seen that with this scaling,
\begin{align}
\cD_\alpha {W} &=\partial_\alpha {W}^{(1)}_{\mathrm{susy}} + \frac{m_{\Lambda}^{r-3}}{\mpl^r} W^{(2)}_{\mathrm{susy}} \partial_\alpha
{W}^{(1)}_{\mathrm{susy}}\non
 &\phantom{=}+ \frac{\partial_\alpha
{K^{(1)}}}{\mpl^2}
\left(m_{\Lambda}^{3-r} \mpl^r +
W_{\mathrm{susy}}^{(1)}+W_{\mathrm{susy}}^{(2)}+
\frac{m_{\Lambda}^{r-3}}{\mpl^r}W_{\mathrm{susy}}^{(1)}W_{\mathrm{susy}}^{(2)}
\right)\to\partial_\alpha W_{\mathrm{susy}}^{(1)} \;,
\end{align}
in the limit $\mpl\rar \infty$ if and only if $0<r<2$
and thus
\begin{align}
\label{eq:1}
V_{\mathrm{sugra}}= \ee^{K/\mpl^2} \left( |\cD {W}|^2-
\frac{3|W|^2}{\mpl^2}\right)\to\sum_j|\partial_j
W_{\mathrm{susy}}^{(j)}|^2-3m^{2(3-r)}_\Lambda\mpl^{2(r-1)} + \cO \left( \frac{1}{\mpl} \right)\;.
\end{align}
For $r< 1$ the manifestly constant term in the potential vanishes
as well and we recover the strict decoupled field theory result, with the gravitino mass going to zero as $m_{3/2}=\langle W \rangle\mpl^{-2}=m_{\Lambda}^{3-r}\mpl^{r-2} = \frac{m_{\mathrm{susy}}^2}{\sqrt{3} \mpl}$.\footnote{This also ensures that the mixing between the two sectors in the fermion mass matrix $m_{i\alpha}= \ee^{G/2} \left( \nabla_i G_{\alpha} + G_i G_{\alpha} \right)$ does not interfere with the decoupling, since it is proportional to the gravitino mass and thus vanishes naturally.} We see that the gravitino mass is independent of $r$ in physical scales.

The parameter $r$ should not be larger than unity for the new
decoupling limit to be well defined. For the special case $r=1$ \cite{Cremmer:1982vy},
the potential has an additional overall ``cosmological'' constant.
For a generic non-gravitational field theory in which $\mpl \rar
\infty$ this is just an overall shift of the potential, which we
can arbitrarily remove since it does not change the physics.
Nevertheless from a formal point of view, we know that absolute
ground state energy of a globally supersymmetric theory equals
zero, as a result of the supersymmetry algebra $\{Q,Q\}=H$. For
this reason it is more natural to restrict the value of $r$ to the
range $0<r<1$.

\section{Concluding remarks}
Let us conclude with a comment on the physical meaning behind the
scaling \eqref{eq:scaling}. It {may} appear that we have changed
the canonical RG-scaling of the theory. This is not quite true.
For the interacting terms in the potential, it is the coefficients
in the product
$W_0^{(2)}W_{\mathrm{dyn}}^{(1)}=W_{\mathrm{susy}}^{(1)}$ that
ought to obey canonical RG-scaling. This precisely corresponds to
holding $W_{\mathrm{susy}}^{(j)}$ fixed as $\mpl \rar \infty$ (see
footnote \ref{fn:1}). On the other hand, the scaling of the
constant term in the potential has changed from its canonical
value. However, this is very natural in a supersymmetric theory.
The constant term, $\prod_j W_0^{(j)}$, equals the ground state
energy. Precisely supersymmetric theories can ``naturally''
explain non-canonical scaling of the cosmological constant (at the
loop level; the scaling of the bare ground state energy can be
different in every model). A non-integer power is strange but
$r=1$ is certainly a viable option in a supersymmetry-breaking
ground state: it is the natural scaling in theories with higher
supersymmetry
\cite{Ferrara:1994kg} when combined with a subleading
$\log(\mpl/m_{\mathrm{susy}})$ breaking. Our engineering analysis
only focuses on power-law scaling and these can always have
subleading logarithms. ($r=2$ would correspond to the cosmological
constant for a spontaneously broken ${\cal N}=1$ theory due to
mass splitting).

Finally, the novel scaling in \eqref{eq:scaling} can be readily
generalized to an arbitrary number of sectors. For $s$ sectors,
writing $W^{(j)}=W^{(j)}_0+W^{(j)}_{\mathrm{dyn}}$,
\begin{align}
W&=\frac{1}{\mpl^{3(s-1)}}\prod_{j=1}^sW^{(j)}\nonumber\\
&=\frac{1}{\mpl^{3(s-1)}}\left[\prod_{j=1}^sW_0^{(j)}+\sum_{k=1}^s\left(W_{\mathrm{dyn}}^{(k)}\prod_{j\neq
k}^sW_0^{(j)}\right)+\sum_{l>k}^s\left(W_{\mathrm{dyn}}^{(k)}W_{\mathrm{dyn}}^{(l)}\prod_{j\neq
k,l}^sW_0^{(j)}\right)+\ldots\right]\;.\label{eq:multisectorW}
\end{align}
We want the last and all further terms to scale away as $\mpl^{-r}$ and higher with $r>0$, while the second
term(s) should be constant. As a consequence the first term will scale as $\mpl^{r}$. Assuming a scaling that is homogeneous
across sectors, this implies
\begin{equation}\label{eq:multisectorscaling}
W_0^{(j)}\sim \mpl^{\frac{3(s-1)+r}{s}}\;, \qquad
W_{\mathrm{dyn}}^{(j)}\sim \mpl^{\frac{(3-r)(s-1)}{s}} \;,
\end{equation}
for each of the $j\in\{1,\ldots,s\}$. With this scaling, a general
term consisting of $l$ dynamical superpotentials and $s-l$
constant parts, scales as
\begin{equation}\label{eq:scalingPlancksuppressed}
\frac{W_{\mathrm{dyn}}^lW_0^{s-l}}{\mpl^{3(s-1)}}\sim \mpl^{r(1-l)} \;,
\end{equation}
and as constructed any term containing dynamical interactions between
sectors, $l>2$, is Planck-suppressed. To ensure a vanishing constant term as in eq. \eqref{eq:1}, $r$ is again limited to the range $0<r<1$.

\acknowledgments
We thank G.~Palma, M.~Postma, S.~Vandoren and A.~Westphal for useful comments and discussions.
This research was supported in part by a VIDI Innovative Research
Incentive Grant (K. Schalm) from the
Netherlands Organisation for Scientific Research (NWO), a VICI Award
(A. Ach\'ucarro) from the Netherlands Organisation for Scientific
Research (NWO) and the Dutch Foundation for Fundamental Research on
Matter (FOM).


\providecommand{\href}[2]{#2}\begingroup\raggedright\endgroup

\end{document}